\documentclass[11pt]{JHEP3}
\usepackage{amssymb}
\usepackage{graphicx}

\newcounter{saveeqn}
\newcommand{\alpheqn}{\setcounter{saveeqn}{\value{equation}}%
 \stepcounter{saveeqn}\setcounter{equation}{0}%
 \renewcommand{\theequation}
     {\thesection.\mbox{\arabic{saveeqn}\alph{equation}}}}
\newcommand{\reseteqn}{\setcounter{equation}{\value{saveeqn}}%
  \renewcommand{\theequation}{\thesection.\arabic{equation}}}

\title{First Order Born-Infeld Hydrodynamics via Gauge/Gravity Duality}

\author{H. S. Tan\\
    Berkeley Center for Theoretical Physics and Department of Physics,\\
	University of California, Berkeley, CA 94720-7300\\
	E-mail: \email{haisiong\_tan@berkeley.edu}}

\received{\today} 		
\accepted{\today}		

\abstract{By performing a derivative expansion on a class of boosted Born-Infeld-$\textrm{AdS}_{5}$ black branes, we study the hydrodynamics of the dual field theory - in the spirit of AdS/CFT correspondence. We determine the fluid dynamical stress-energy tensor to first order, and find that the ratio of the shear viscosity to entropy density conforms to the universal value of $1/4\pi$ to all orders of the inverse of the Born-Infeld parameter.}

\newcommand{\be}{\begin{equation}
\addtolength{\abovedisplayskip}{\extraspaces}
\addtolength{\belowdisplayskip}{\extraspaces}
\addtolength{\abovedisplayshortskip}{\extraspace}
\addtolength{\belowdisplayshortskip}{\extraspace}}
\newcommand{\ee}{\end{equation}}

\newcommand{\ba}{\begin{eqnarray}
\addtolength{\abovedisplayskip}{\extraspaces}
\addtolength{\belowdisplayskip}{\extraspaces}
\addtolength{\abovedisplayshortskip}{\extraspace}
\addtolength{\belowdisplayshortskip}{\extraspace}}
\newcommand{\ea}{\end{eqnarray}}

\begin{document}

\section{Introduction}

The AdS/CFT correspondence \cite{Maldacena} has clearly shown its prowess in elucidating various aspects of strongly coupled gauge theories. One of the first examples encountered in string theory was the study of $\mathcal{N}=4\,\,\,SU(N)$ supersymmetric-Yang-Mills theory on $R^{3,1}$ with a nonzero chemical potential. For this theory, which is dual to type IIB supergravity on $AdS_{5} \times S^{5}$, the ratio of the shear viscosity $\eta$ to the entropy density $s$ has been computed via the Kubo formula \cite{Policastro}, as well as a recently developed technique of mapping a hydrodynamic expansion of the boundary to a gradient expansion in the bulk \cite{Minwalla}. In the large $N$ limit and at strong t'Hooft coupling, $\eta/s = 1/4\pi$. Indeed by now, we know that this lower bound enjoys a well-known universality of being applicable to an enormous class of gauge theories with holographic gravity duals. However, there have also been many insightful attempts to derive plausible corrections to this ratio. For example, in \cite{Myers}, higher derivative terms with the five-form Ramond-Ramond flux were taken carefully into account; while in \cite{Shenker}, motivated by the vast stringy landscape, deviations from this ratio were found when the field theory is mapped to gravity with Riemann curvature square terms.

Recently, an interesting effort was made in \cite{Cai} in checking this ratio for the field theory with Einstein-Born-Infeld (EBI) gravity as the dual. First, let us briefly discuss the relevance of EBI gravity. Now, although it is well-known that the Dirac-Born-Infeld (DBI) action governs the low energy dynamics of D-branes \cite{Fradkin}, it is not clear that the EBI action (see (\ref{action})) can arise from some consistent truncation of any type IIB supergravity. Yet arguably, the EBI Lagrangian is the simplest nonlinear generalization of the U(1) gauge field in Einstein-Maxwell-AdS theory to which it reduces in the infinite limit of the Born-Infeld parameter $\beta = 1/2\pi \alpha$, where $\alpha$ is the string tension . Another good reason to study EBI-dual field theories is that the scaling limit which preserves the supergravity solutions does give the entire DBI action on the gauge theory side \cite{Alwis}. Under a dimensional reduction coupled with the Monge/static gauge, Born-Infeld terms should then arise in the Lagrangian \cite{Gibbons}. In the context of gauge/gravity duality, this motivates one to use the action in (\ref{action}) to construct and study features of a Born-Infeld-corrected gauge theory.

In \cite{Cai}, the $\eta/s$ ratio was beautifully checked to be actually still $1/4\pi$ up to the first order of the inverse square of $\beta$, by virtue of the Kubo formula. The authors in \cite{Cai} interpreted this result as essentially saying that the universality of the ratio $1/4\pi$ is simply because the different gauge field theories are dual to the same Einstein-Hilbert gravity. Indeed, this perspective is supported by other works such as in \cite{Brustein}. However, a caveat is that the calculation in \cite{Cai} holds only to the first non-trivial order of $1/\beta^{2}$. This difficulty can be easily traced to the fact that the equations of motion of EBI gravity are rather complex. Recall that in using Kubo formula, one perturbs the graviton to compute Green's functions and then solve the resulting coupled differential equations for the metric perturbations. The shear viscosity is then picked up from the low frequency behavior of the perturbation fields. Unfortunately, the form of EBI gravitational action appears to make such a calculation analytically impossible if one demands a conclusion for $\eta/s$ to all orders of inverse $\beta$.

In this brief paper, we reconsider this problem from the viewpoint of the recently developed technique in \cite{Minwalla} which relates the hydrodynamic regime of the gauge theory to black holes that are asymptotically $\textrm{AdS}_{5}$ via a derivative expansion. The shear viscosity comes from a first-order (in the derivative expansion) dissipative correction to the fluid's stress-energy tensor which can be interpreted as the stress tensor of the dual boundary field theory in its de-confined phase. We find that this different method actually yields $\eta/s = 1/4\pi$ to all orders of inverse $\beta$, thus strengthening the result in \cite{Cai} where the Kubo formula was used instead.

Now the main idea behind the technique in \cite{Minwalla} can be briefly summarized as follows: begin with a $d$ parameter set of exact asymptotically $\textrm{AdS}_{d+1}$ black branes parameterized by constant temperature and velocities. The Goldstone philosophy is invoked to promote these temperature and velocities to slowly-varying coordinate-dependent fields. The Einstein's equations are then solved perturbatively in the number of field theory derivatives, order by order in the derivative expansion in $1/LT$, where $L$ is the length scale of variations and $T$ the Hawking temperature. The first few terms in the expansion turn out to be familiar terms in hydrodynamics, with the zeroth order terms equal to the stress tensor of an ideal fluid and the coefficient of its first-order term being its shear viscosity. The higher-order terms describe other dissipative effects, and from which we can read off quantities like relaxation times. In essence, this is the so-called fluid/gravity correspondence (see \cite{Ambrosetti} for a nice review).

Since our black branes reduce to those recoverable from Einstein-Maxwell-AdS gravity in the infinite $\beta$ limit, it is appropriate to summarize what has been achieved in literature so far in the latter. Independently in \cite{Hur}, \cite{Banerjee} and \cite{Erdmenger}, electrically charged $\textrm{AdS}_{5}$ black branes are studied via this fluid/gravity correspondence. Now, there is a charge current which is varying as well. The conserved charge of the CFT (besides energy-momentum) is just the number of particles that constitute the fluid. First- and second-order transport coefficients (including $\eta/s$) have been calculated too. It is also worthwhile to note that the treatment in \cite{Banerjee} and \cite{Erdmenger} contains an additional Chern-Simons term which renders the gravity action to be a consistent truncation of type IIB supergravity on $AdS_{5} \times S^{5}$ (see \cite{Emparan1} for the KK reduction ansatz).

Our paper is organized as follows: in section 2, we first review the known black hole solutions to the EBI action, boost them via a set of constant 4-velocity parameters, and choose the chart of Eddington-Finkelstein coordinates. In section 3, we carry out the derivative expansion and solve for the global metric and gauge field at first order. With these, we will read off the ratio of $\eta/s$ to be $1/4\pi$ exact to all orders of inverse $\beta$. In section 4, we end off with discussions about future directions. We shall use naturalized units in which $\hbar = c = 1$, with the $5D$ Newtonian gravitational constant $G=1/16\pi$.

\section{Boosted Born-Infeld-$\textrm{AdS}_{5}$ black branes}

We begin with the action
\begin{equation}
\label{action}
S = \int_{\mathcal{M}} \textrm{d}^{5}x \sqrt{-g} \Bigg( R - 2\Lambda + \frac{\beta^{2}}{g^{2}} \Bigg[1-\sqrt{1+\frac{F^{2}}{2\beta^{2}}}\,\, \Bigg] \Bigg) + 2 \int_{\mathcal{\partial M}} \textrm{d}^{4}x \sqrt{-\gamma} \Bigg( K + \sqrt{\frac{3\Lambda}{2}}\bigg(1-\frac{R^{(4)}}{2\Lambda} \bigg) \Bigg)
\end{equation}
where $R,\, R^{(4)}$ are the Ricci scalars of the bulk and boundary metric respectively, $\Lambda$ being the cosmological constant, $\gamma$ the boundary metric, $g$ the $R-$charge coupling, and $K$ the trace of the extrinsic spacetime curvature $K^{\mu\nu} = -\frac{1}{2} (\nabla^{\mu}n^{\nu} + \nabla^{\nu}n^{\mu})$, with $n^{\mu}$ being the outward pointing normal vector to the boundary. The action in (\ref{action}) consists of the usual terms in EBI gravity, with $K$ being the Gibbons-Hawking term and the term thereafter being the counter-terms which have to be present to cancel divergences due to the infinite volume of $\textrm{AdS}_{5}$ (see \cite{Kraus} for the original derivation). Now, the equations of motion to (\ref{action}) read
\begin{eqnarray}
\label{e.o.m.1}
R_{\mu\nu} - \frac{1}{2}Rg_{\mu\nu} + \Lambda g_{\mu\nu} &=& \frac{1}{2g^{2}}\Bigg(\frac{F_{\mu\alpha}{F_{\nu}}^{\alpha}}{\sqrt{1+F^{2}/2\beta^{2}}} + \beta^{2}g_{\mu\nu}\bigg(1-\sqrt{1+F^{2}/2\beta^{2}}\bigg) \Bigg)\\
\label{e.o.m.2}
\nabla_{\mu} \Bigg( \frac{F^{\mu\nu}}{\sqrt{1+F^{2}/2\beta^{2}}}\Bigg) &=& 0
\end{eqnarray}
Exact solutions to (\ref{e.o.m.1}) and (\ref{e.o.m.2}) were first found in \cite{Pang} and \cite{Dey}, and the black holes were topological in nature. Since the dual CFT resides in $\Re^{1,3}$, we require the event horizon to be Ricci-flat. Further, to make the horizon apparently regular, we work in incoming Eddington-Finkelstein coordinates. Our black brane metric reads (we set $\Lambda = -6$ henceforth)
\begin{equation}
\label{metric0}
\textrm{d}s^{2}=-r^{2}f(r)\textrm{d}v^{2} + 2\textrm{d}r\textrm{d}v + r^{2} \big( \textrm{d}x_1^2 + \textrm{d}x_2^2 + \textrm{d}x_3^2 \big)
\end{equation}
where
\begin{equation}
\label{V}
f(r) = \bigg( 1 - \frac{2m}{r^{4}} + \frac{q^{2}}{r^{6}} \bigg\{ \frac{3}{2} {}_{2}F_{1} \bigg[\frac{1}{3},\frac{1}{2}, \frac{4}{3}, \frac{-12g^{2}q^{2}}{\beta^{2}r^{6}} \bigg] \bigg\} \bigg) + \frac{\beta^{2}}{12g^{2}} \Bigg( 1 - \sqrt{1 + \frac{12g^{2}q^{2}}{\beta^{2}r^{6}} } \Bigg)
\end{equation}
with ${}_{2}F_{1}$ being the hypergeometric function that admits a convergent series expansion for large $\beta$ or $r$. The isometry of (\ref{metric0}) allows us to perform a boost with a constant time-like Minkowski 4-vector $u^{\mu}$, with $u^{2}=-1$. Our final metric thus reads
\begin{equation}
\label{metric}
\textrm{d}s^{2}=-r^{2}f(r)\big[u_{\mu}\textrm{d}x^{\mu} \big]^{2} - 2u_{\mu}\textrm{d}x^{\mu}\textrm{d}r + r^{2} P_{\mu \nu} \textrm{d}x^{\mu} \textrm{d}x^{\nu}
\end{equation}
where
\begin{equation}
\label{boost}
u^{0}=\frac{1}{\sqrt{1-b_{i}^{2}}},\,\,\,u^{i} = \frac{b_{i}}{\sqrt{1-b_{i}^{2}}},\,\,\,P_{\mu\nu}= \eta_{\mu\nu}+u_{\mu}u_{\nu}
\end{equation}
while the U(1) gauge forms are
\begin{equation}
\label{gauge}
F=-g\frac{2\sqrt{3}\beta q}{\sqrt{\beta^{2}r^{6}+12g^{2}q^{2}}}u_{\mu} \textrm{d}x^{\mu} \wedge \textrm{d}r,\,\,\,\,A=\bigg( \frac{\sqrt{3}gq}{r^{2}}{}_{2}F_{1} \bigg[\frac{1}{3},\frac{1}{2}, \frac{4}{3}, \frac{-12g^{2}q^{2}}{\beta^{2}r^{6}} \bigg] u_{\mu} + eA_{\mu}^{\rm{ext}} \bigg) \textrm{d}x^{\mu}
\end{equation}
Following \cite{Hur}, we have purposefully distinguished between the R-charge coupling $g$ and external electric charge coupling $e$. The Hawking temperature $T$ and chemical potential $\mu$ of the black branes depend on the outer event horizon radius $r_{+}$ because
\begin{eqnarray}
\label{temp}
T&=&\frac{1}{\pi r_{+}} \Bigg( r_{+}^{2} + \frac{\beta^{2}r_{+}^{2}}{12g^{2}} - \frac{\beta^{2}r_{+}^{2}}{12g^{2}} \bigg( 1 + \frac{12g^{2}q^{2}}{\beta^{2} r_{+}^{6}} \bigg)^{1/2}  \Bigg) \\
\label{chem}
\mu&=&\frac{\sqrt{3}gq}{r_{+}^{2}}{}_{2}F_{1} \bigg[\frac{1}{3},\frac{1}{2}, \frac{4}{3}, \frac{-12g^{2}q^{2}}{\beta^{2}r_{+}^{6}} \bigg]
\end{eqnarray}
The constants parameterizing our black branes form the set $\{m, q, A^{\rm{ext}}, b\}$. We now proceed to perform the derivative expansion of fluid/gravity correspondence, i.e. we perturb the bulk metric to evoke the hydrodynamical degrees of freedom of the boundary's CFT.

\section{The derivative expansion to first order}

The first step is to uplift the constant parameters to become slowly-varying functions of the transverse coordinates $x^{\mu} \equiv (v, x_{i})$. Since $\{m, q, A^{\rm{ext}}, b\}$ are no longer constants, the metric (\ref{metric}) and gauge fields (\ref{gauge}) are no longer valid solutions to (\ref{e.o.m.1}) and (\ref{e.o.m.2}). We shall extend this exact solution by adding $n$-th order metric components to (\ref{metric}) order by order in a derivative expansion. To set the stage right, we decompose the $n$-correction metric into\footnote{Our convention mainly follows that of \cite{Hur}.}
\begin{equation}
\label{correctedmetric}
\textrm{d}s^{(n)^{2}} = \frac{k^{(n)}}{r^{2}} \textrm{d}v^{2} + 2h^{(n)} \textrm{d}v \textrm{d}r + 2 \frac{j^{(n)}_{i}}{r^{2}} \textrm{d}v \textrm{d}x^{i} + r^2 \Big(\alpha^{(n)}_{ij} - \frac{2}{3} h^{(n)}\delta_{ij} \Big) \textrm{d}x^{i} \textrm{d}x^{j}
\end{equation}
and similarly, the gauge fields into
\begin{equation}
\label{correctedgauge}
A^{(n)} = a_{v}^{(n)} (r) \textrm{d}v + a_{i}^{(n)} (r) \textrm{d}x^{i}
\end{equation}
where we have adopted an axial gauge for the gauge field. It will turn out that because the background metric (\ref{metric}) preserves a spatial $SO(3)$ symmetry, this symmetry allows us to solve separately for the $SO(3)$ scalars \{ $k^{(n)},\,h^{(n)}$ \}, the $SO(3)$ vectors $j^{(n)}_{i}$ and the $SO(3)$ symmetric traceless two-tensor $\alpha^{(n)}_{ij}$. As mentioned in \cite{Minwalla}, this is crucial for the integrability of the system. Following the methodology in \cite{Minwalla}, we define the tensors:

\begin{equation}
\label{e.o.m.3}
W_{AB}=R_{AB} + 4g_{AB} + \frac{1}{2g^{2}}\Bigg(       \frac{F_{AM}{F^{M}}_{B}}{\sqrt{1+F^{2}/2\beta^{2}}} + \frac{1}{6} g_{AB}\Bigg[ \frac{2F^{2}}{\sqrt{1+\frac{F^{2}}{2\beta^{2}}}} +     4\beta^{2} \Bigg(1-\sqrt{1+\frac{F^{2}}{2\beta^{2}}}\Bigg)
\Bigg]           \Bigg) \\
\end{equation}
\begin{equation}
\label{e.o.m.4}
W_{A}=\nabla_{B} \Bigg( \frac{{F^{B}}_{A}}{\sqrt{1+F^{2}/2\beta^{2}}}\Bigg)
\end{equation}
which vanish when the equations of motion are satisfied. Now consider these tensors in the neighborhood of a point $x_{0}^{\mu}=0$ (but at arbitrary $r$) with the choice of $u^{\mu}=(1,0,0,0), b=b_{0}, m=m_{0}, q=q_{0}, A^{\rm{ext}}=A^{\rm{ext}}_{0}$. When we uplift $\{m, q, A^{\rm{ext}}, b\}$ to be functions of $x^{\mu}$, and carry out a derivative expansion, the extra terms that emerge from (\ref{e.o.m.3}) and (\ref{e.o.m.4}), which we denote as $S_{A}, S_{AB}$, are proportional to the derivatives of the parameter functions . To ensure that the tensors $W_{AB}, W_{A}$ vanish, we have to add the correction terms of (\ref{correctedmetric}) and (\ref{correctedgauge}) to the metric (\ref{metric}).

These imply that we can obtain the metric to first order in this derivative expansion by solving coupled differential equations involving the various correction terms and first-order derivatives of $\{m, q, A^{\rm{ext}}, b\}$. Now once the solution is obtained around $x_{0}^{\mu}$, it can be uniquely extended to the rest of the manifold. This process can, in principle, be repeated iteratively, with complication coming from the increasing complexity of the source terms $S_{AB}, S_{A}$. Assuming that we have obtained the $(n-1)^{\rm{th}}$ order solution, we can calculate the $n^{\rm{th}}$ derivative order source terms and obtain $n^{\rm{th}}$ order correction terms in metric and gauge fields. We also expect constraint equations which should lead nicely to conservation laws of the hydrodynamical theory at the boundary. Indeed, we will verify this explicitly later.

In implementing this technique for (\ref{metric}), we find the following equations and definitions useful:
\begin{eqnarray}
\label{definitions}
&\delta R_{\mu\nu} \approx \nabla_{a}\nabla_{(\mu}{h_{\nu)}}^{a} - \frac{1}{2}\nabla_{\mu}\partial_{\nu}h - \frac{1}{2}\nabla^{2}h_{\mu\nu},\,\,(h \equiv \delta g)\nonumber
\cr
\cr
&\frac{q}{r^{6}}{}_{2}F_{1}\big(\frac{1}{3}, \frac{1}{2}, \frac{4}{3}, \frac{-12g^{2}q^{2}}{\beta^{2}r^{6}} \big)-
\frac{3q}{8r^{6}}\big(\frac{12g^{2}q^{2}}{\beta^{2}r^{6}}\big){}_{2}F_{1}\big(\frac{4}{3}, \frac{3}{2}, \frac{7}{3}, \frac{-12g^{2}q^{2}}{\beta^{2}r^{6}} \big)-\frac{q}{r^{6}}\big(1+\frac{12g^{2}q^{2}}{\beta^{2}r^{6}}\big)^{-1/2}=0\nonumber
\cr
\cr
&H \equiv \frac{gr^{2}}{\sqrt{3}}\partial_{r} \big(\frac{r^{3}}{2q}\Xi \big),\,\, \Xi \equiv \frac{2q}{r^{6}}{}_{2}F_{1}\big(\frac{1}{3}, \frac{1}{2}, \frac{4}{3}, \frac{-12g^{2}q^{2}}{\beta^{2}r^{6}} \big),\,\, \Upsilon \equiv \frac{2\sqrt{3}\beta g q }{\sqrt{\beta^{2}r^{6}+12g^{2}q^{2}}},\nonumber
\cr
\cr
&\Theta \equiv \big(1-\frac{\Upsilon^{2}}{\beta^{2}} \big)^{-1/2},\,\, G_{1} \equiv -\frac{12g^{2}q^{2}}{\beta^{2}r^{6}}\Theta\nonumber
\\
\end{eqnarray}
With the aid of (\ref{definitions}) and after some algebra, we obtain the following equations for $W_{A}$ \footnote{Note that we have suppressed the superscript label $(n)$ in the equations which should be valid for any order in the derivative expansion. In each order, only the expressions for $S$ are different.}:
\begin{eqnarray}
\label{m1}
W_{i}&=&\frac{1}{r}\partial_{r} \Bigg(\Theta \bigg(r^{3}f\partial_{r}a_{i}-\frac{j_{i}\Upsilon}{r}\bigg)\Bigg) - S_{i}=0\,,\\
\label{m2}
W_{r}&=&-\frac{1}{r^{3}} \partial_{r}\Bigg( r^{3} \partial_{r}a_{v} \big(1-G_{1} \big) + 2hr^{3}\Upsilon \big(1 - G_{1}/2 \big) \Bigg) - S_{r}=0\,,\\
\label{m3}
W_{v}&=&\frac{f}{r} \partial_{r}\Bigg( r^{3} \partial_{r}a_{v} \big(1-G_{1} \big) + 2hr^{3}\Upsilon \big(1 - G_{1}/2 \big) \Bigg) - S_{v}=0\,,\\
\nonumber
\end{eqnarray}
From (\ref{m2}) and (\ref{m3}), we observe the constraint:
\begin{equation}
\label{maxwellconstraint}
S_{v} + r^{2}fS_{r}=0
\end{equation}
Similarly, the tensors of (\ref{e.o.m.3}) were found to simplify to:
\begin{eqnarray}
\label{E1}
&&W_{vv}=f\Bigg\{ \frac{2r^2 \Upsilon \Theta}{3g^{2}} \left(1- \frac{G_{1}}{4\Theta} \right) \partial_{r}a_{v} - 8r^{2} \left( 1 - \frac{\Upsilon^{2}\Theta}{24g^{2}} \left(1 - \frac{G_{1}}{2\Theta} - \frac{2\Theta(1-\Theta)}{G_{1}} \right) \right)h\cr
&&\qquad \qquad- r^{2} \partial_{r} \left(r^{2}f \right) \partial_{r}h - \frac{r}{2}\partial_{r} \left(\frac{\partial_{r} k}{r}   \right) \Bigg\}- S_{vv} \,,\cr
\nonumber
\cr
&&W_{vr}=\frac{2\Upsilon \Theta}{3g^{2}} \left(1- \frac{G_{1}}{4\Theta} \right) \partial_{r}a_{v} - 8 \left( 1 - \frac{\Upsilon^{2}\Theta}{24g^{2}} \left(1 - \frac{G_{1}}{2\Theta} - \frac{2\Theta(1-\Theta)}{G_{1}} \right) \right)h \cr
&&\qquad \qquad -\partial_{r} \left(r^{2}f \right) \partial_{r}h - \frac{1}{2r}\partial_{r} \left(\frac{\partial_{r} k}{r}   \right) - S_{vr}\,,\cr
\nonumber
\cr
&&W_{vi}=f\Bigg\{\frac{r^{2}\Upsilon\Theta\partial_{r}a_{i}}{2g^{2}} - \frac{r^{3}}{2} \partial_{r} \left( \frac{\partial_{r} j_{i}}{r^{3}}   \right) \Bigg\} - S_{vi}\,,\cr
\nonumber
\cr
&&W_{rr}=\frac{1}{r^{5}} \partial_{r} \left( r^{5} \partial_{r} h \right) - S_{rr}\,,\cr
\nonumber
\cr
&&W_{ri}=-\frac{\Upsilon\Theta \partial_{r} a_{i}}{2g^{2}} + \frac{r}{2} \partial_{r} \left( \frac{\partial_{r}j_{i}}{r^{3}} \right) - S_{ri}\,,\cr
\nonumber
\cr
&&W_{ij}\delta^{ij}= \frac{r^{2}\Upsilon (\Theta - G_{1})}{g^{2}} \partial_{r}a_{v} + 24r^{2} \left( 1 - \frac{\Upsilon^{2}\Theta}{24g^{2}} \left(1 + \frac{G_{1}}{\Theta} - \frac{2\Theta(1-\Theta)}{G_{1}} \right) \right)h\cr
&&\qquad \qquad + \frac{\partial_{r}(r^{11}f\partial_{r}h)}{r^{7}} + \frac{3 \partial_{r}k}{r}- S_{ij}\delta^{ij}\,,\cr
\nonumber
\cr
&&W_{ij}= \frac{1}{3}\delta_{ij}\left(\delta^{kl}W_{kl}\right) - \frac{1}{2r} \partial_{r} \left(r^{5}f \partial_{r} \alpha_{ij}\right) - S_{ij} + \frac{1}{3} \delta_{ij} \left(\delta^{kl}S_{kl}\right)\,,\cr\nonumber
\end{eqnarray}
\begin{equation}
\label{E}
\end{equation}
These tensors of (\ref{E1}) vanish when (\ref{e.o.m.1}) is satisfied. The analogue of (\ref{maxwellconstraint}) reads:
\alpheqn
\begin{eqnarray}
\label{einsteinconstraint}
S_{vv} + r^{2}fS_{vr} = 0\,,\\
S_{vi} + r^{2}fS_{ri} = 0\,,
\end{eqnarray}
\reseteqn
We can do some number counting as a consistency check: there are 15 equations with 5 constraints to solve for the 10 unknown metric components in (\ref{correctedmetric}), and 5 equations with 1 constraint to solve for the 4 unknown gauge fields in (\ref{correctedgauge}). Let us now proceed to list down the first order source terms which we find to be:
\\
\begin{eqnarray}
\label{source}
&&S_{vv}=-\frac{1}{2}\partial_{r} \big( r^{2}f \big) \partial_{i} b_{i} - \frac{3}{r^{3}} \partial_{v}m + \frac{3}{2} r \Xi \partial_{v}q \,,\cr
\nonumber
\cr
&&S_{vr}=\frac{1}{r}\partial_{i} b_{i}\,,\cr
\nonumber
\cr
&&S_{vi}=\bigg(\frac{4m}{r^{3}} + \frac{3rf}{2} \bigg) \partial_{v} b_{i} + \frac{\partial_{i}m}{r^{3}} - \frac{\sqrt{3}q e}{r^{3}g} F_{vi}^{\rm{ext}}\,,\cr
\nonumber
\cr
&&S_{rr}=S_{r}=0\,,\cr
\nonumber
\cr
&&S_{ri}=-\frac{3}{2r} \partial_{v} b_{i} \,,\cr
\nonumber
\cr
&&S_{ij}=r\bigg\{ \delta_{ij} \partial_{k} b_{k} + \frac{3}{2} \big( \partial_{i} b_{j} + \partial_{j} b_{i} \big) \bigg\} \,,\cr
\nonumber
\cr
&&S_{v}= g\frac{2\sqrt{3}g q}{r^{3}} \big(\partial_{v} q + q \partial_{i} b_{i} \big)\,,\cr
\nonumber
\cr
&&S_{i}=-\frac{\Theta H}{r} \bigg[ 1 + r\frac{\partial_{r} \Theta}{\Theta} + \frac{r\Upsilon}{q H \Theta^{2}} \bigg] \partial_{i}q + \frac{1}{r} \partial_{r} \bigg[ \big(r\Theta \big) \frac{\sqrt{3}g q }{r^{2}} {}_{2}F_{1} \bigg] \partial_{v} b_{i} - \frac{1}{r} e F_{vi}^{\rm{ext}} \partial_{r} \big( \Theta r \big) \,,\cr
\\
\end{eqnarray}
In particular, these source terms imply the conservation equations:
\begin{eqnarray}
\label{conservation0}
4m\partial_{i}b_{i} + 3\partial_{v}m &=& 0\,,\cr
\partial_{v}q + q \partial_{i}b_{i} &=& 0\,,\cr
\partial_{i}m + 4m \partial_{v}b_{i} &=& \sqrt{3}\frac{e q}{g} F_{vi}^{\rm{ext}}
\end{eqnarray}
Later, we will see that (\ref{conservation0}) can be expressed covariantly in terms of the energy-momentum and charge current tensors. We now proceed to solve (\ref{m1})-(\ref{m2}) and (\ref{E1}) partially, first withholding the substitution of the source term expressions in (\ref{source}) since these results can be useful for higher-order treatments. Comparing with the analysis in \cite{Hur}, \cite{Banerjee} and \cite{Erdmenger} which were concerned with the Einstein-Maxwell-AdS limiting case, we find that our solutions are very similar, and the complication only arises in the solving for $a_{i}$ and $j_{i}$. Indeed, it is straightforward to derive
\\
\begin{eqnarray}
\label{partialsolution}
a_{v}&=&-\int^{r}\frac{2-G_{1}(y)}{2(1-G_{1}(y))}2h(y)\Upsilon(y) \textrm{d}y - \int^{r} \frac{1}{y^{3}(1 - G_{1}(y))} \int^{y} x^{3} S_{r}(x) \textrm{d}x\,\textrm{d}y\,,\cr
k&=&\int^{r} \frac{x}{3} \Bigg\{ -\frac{x^{2}\Upsilon}{g^{2}} \big(\Theta - G_{1} \big)\partial_{x}a_{v} - \frac{1}{x^{7}} \partial_{x} \big( x^{11}f\partial_{x}h \big) + S_{ij}\delta^{ij}\cr
&&- 24x^{2} \bigg\{1 - \frac{\Upsilon^{2} \Theta}{24g^{2}} \bigg(1+ \frac{G_{1}}{\Theta} - \frac{2\Theta(1-\Theta)}{G_{1}} \bigg) \bigg\}h\Bigg\} \textrm{d}x\,,\cr
h&=&\int^{r} \textrm{d}y \frac{1}{y^{5}} \int^{y} x^{5} S_{rr} (x) \textrm{d}x \,,\cr
\alpha_{ij}&=&-\int^{r} \textrm{d}y \frac{1}{y^{5}f(y)} \int^{y} \textrm{d}x 2x \bigg( S_{ij}(x) - \frac{1}{3} \delta_{ij} \delta^{kl} S_{kl} \bigg)
\end{eqnarray}
and with a set of coupled differential equations for $a_{i}$ and $j_{i}$:
\alpheqn
\begin{eqnarray}
\label{coupled}
-\frac{\sqrt{3}q}{gr^{3}}\Theta \partial_{r}a_{i} + \frac{r}{2} \partial_{r} \bigg( \frac{\partial_{r}j_{i}}{r^{3}}    \bigg) = S_{ri}\,,\\
\partial_{r} \bigg\{ \bigg( r^{3}f\partial_{r} a_{i} - \frac{j_{i}\Upsilon}{r} \bigg) \Theta \bigg\} = rS_{i}\,,
\end{eqnarray}
\reseteqn
From the source term expressions in (\ref{source}), we obtain:
\begin{eqnarray}
\label{solution1}
a_{v}=h=0,\qquad k = \frac{2}{3}r^{3} \partial_{i}b_{i}\,,\cr
\alpha_{ij} = \alpha(r) \big\{ \partial_{i} b_{j} + \partial_{j} b_{i} - \frac{2}{3} \delta_{ij} \partial_{k} b_{k} \big\}\,,
\end{eqnarray}
where
\begin{eqnarray}
\label{alpha}
\alpha(r) &=& 3 \int^{r}_{\infty} \textrm{d} t \frac{1}{t^{5}f(t)} \int^{t}_{r_{+}} \textrm{d}s s^{2}\cr
\nonumber
\\
&=&\int^{r}_{\infty} \textrm{d}r \bigg( \frac{1}{r^{2}} - \frac{r^{3}_{+}}{r^{5}} \bigg) \bigg( 1 - \frac{2m}{r^{4}} + \frac{q^{2}}{r^{6}} - \frac{3}{4} \frac{g^{2}q^{4}}{\beta^{2}r^{12}} + \ldots \bigg)^{-1}\cr
\nonumber
\\
&=&\frac{1}{r} - \frac{r^{3}_{+}}{4r^{4}} + \frac{1}{\beta^{2}} \mathcal{O} \big( \frac{1}{r^{5}} \big)
\end{eqnarray}
Note that in (\ref{alpha}) we have expanded $\alpha$ in $\mathcal{O}\big(\frac{1}{r}\big)$ because the AdS/CFT dictionary requires us to read off the coefficient of $1/r^{4}$ in $\alpha$ as we would elaborate later on. To be careful, the $\beta$-correction comes not merely from those in the argument of the hypergeometric functions, but also whenever $m$ occurs. This is because we have taken our perturbation in $\beta$ on the basis of fixed $r_{+}$ and $q$, upon which then $m$ must be expanded in $\beta$ about $m_{0} \equiv \frac{1}{2} \big( r_{+}^{4} + \frac{q^{2}}{r_{+}^{2}} \big)$ - which is simply the mass parameter of the corresponding Reissner-Nordstr\"{o}m-AdS black holes at infinite $\beta$. At this point, it is nice to observe that the $\beta$-correction does not seem to creep into $\alpha$ at the order of $1/r^{4}$. We are thus left with the task of solving (\ref{coupled}). Unfortunately, the eventual expressions are quite complicated, so we will leave details to the Appendix, noting that $j_{i}$ and $a_{i}$ will {\it not} affect our main goal of calculating $\eta/s$\footnote{Their analysis will be needed if we want to calculate transport coefficients such as thermal/electrical conductivities. A simple perturbation scheme in $1/\beta$ is presented in the Appendix which will be useful for this purpose. We leave further applications to future work.}.
At this junction, it suffices to say that even though (\ref{coupled}) cannot be solved analytically (as opposed to the infinite $\beta$ case), one can, as usual, use ordinary perturbation methods to solve them to orders in $1/\beta$. The subtle point is that this has to be accompanied carefully by an expansion in $1/\beta$ in the mass parameter $m$ about $m_{0}$. As shown in the Appendix, our solution (see also \cite{Hur}, \cite{Banerjee} and \cite{Erdmenger}) lies crucially on selecting $r_{+}$ as some of our integration limits. It is then neater to fix $q$ and $r_{+}$, while letting $m=m(\beta)$.

Let us now proceed to discuss the AdS/CFT dictionary relevant to our purpose here. The boundary stress-energy tensor and current can be calculated out of the bulk after an ADM decomposition of the metric:
\begin{equation}
\label{ADM}
\textrm{d}s^{2} = \gamma_{\mu\nu} \big( \textrm{d}x^{\mu} + V^{\mu}\textrm{d}r \big) \big( \textrm{d}x^{\nu} + V^{\nu}\textrm{d}r \big) + N^{2} \textrm{d} r^{2}
\end{equation}
where $\gamma$ is the induced boundary metric. The tensors can then be read off by computing (see, for example, \cite{Ambrosetti}):
\begin{eqnarray}
\label{energy0}
\langle T_{\mu\nu} \rangle &=& \frac{2}{\sqrt{-\gamma}} \frac{\delta S}{\delta \gamma^{\mu\nu}} = 2\big( K_{\mu\nu} - K\gamma_{\mu\nu} - 3\gamma_{\mu\nu} - \frac{1}{2}G_{\mu\nu} \big)
\\
\label{current0}
\langle J^{\mu} \rangle &=& \frac{1}{\sqrt{-\gamma}}\frac{\delta S}{\delta A_{\mu}} = -2 \lim_{r \rightarrow \infty} \frac{r^{2}A^{\mu}}{g^{2}}
\end{eqnarray}
With (\ref{energy0}) and (\ref{current0}), we can obtain the zeroth order boundary current and energy-momentum tensors:
\begin{equation}
T^{\mu\nu}_{(0)} = 2m\big(\eta^{\mu \nu} + 4u^{\mu}u^{\nu} \big)\,,\qquad J^{\mu}_{(0)} = \frac{2\sqrt{3}q}{g}u^{\mu}
\end{equation}
and thus (\ref{conservation0}) can be expressed covariantly as
\begin{equation}
\partial^{\mu} T_{\mu \nu}^{(0)} = 2\sqrt{3}\frac{q e}{g} F_{\mu\nu}^{\rm{ext}}u^{\mu}\,,\qquad \partial^{\mu}J_{\mu}^{(0)} = 0
\end{equation}
To calculate first-order effects, we have to re-write our metric (\ref{metric}) together with the correction terms in a covariant manner. For this purpose, it is useful to define (see for eg., \cite{Hur})
\begin{equation}
j_{i} \equiv j_{b} \partial_{v} b_{i} + j_{q} \big( \partial_{i}q + q \partial_{v}b_{i} \big) + j_{F} F_{vi}^{\rm{ext}}
\end{equation}
\begin{equation}
\sigma^{\mu\nu} \equiv \frac{1}{2}P^{\mu a}P^{\nu b} \big( \partial_{a} u_{b} + \partial_{b} u_{a} \big) - \frac{1}{3} P^{\mu\nu} \partial_{a} u^{a}
\end{equation}
With some manipulation, our first-order metric can now be expressed globally as
\begin{eqnarray}
\label{metricglobal}
\textrm{d}s^{2} &=& -r^{2} f u_{\mu}u_{\nu}\textrm{d}x^{\mu}\textrm{d}x^{\nu} - 2 u_{\mu} \textrm{d}x^{\mu} \textrm{d}r + r^{2}P_{\mu\nu} \textrm{d}x^{\mu} \textrm{d}x^{\nu} + 2r^{2}\alpha\sigma_{\mu\nu} \textrm{d}x^{\mu}\textrm{d}x^{\nu}+ \bigg[ \frac{2r}{3}u_{\mu}u_{\nu}\partial_{w}u^{w}\,\cr \nonumber
\\
&&- \frac{2}{r^{2}} \bigg\{ \frac{1}{2} j_{b} u\cdot\partial(u_{\mu\nu}) + j_{q}u_{\mu}u^{a}[\partial_{a}(qu_{\nu})-\partial_{\nu}(qu_{a})] + j_{F}u^{a}F_{a\nu}^{\rm{ext}} \bigg\} \bigg] \textrm{d}x^{\mu}\textrm{d}x^{\nu}\
\end{eqnarray}
Now, we can then insert (\ref{metricglobal}) into (\ref{energy0}) to find that
\begin{equation}
\label{mainresult}
T_{\mu\nu} = 2m(\eta_{\mu\nu}+4u_{\mu}u_{\nu}) - 2r^{3}_{+} \sigma_{\mu\nu}
\end{equation}
From (\ref{mainresult}), we can read off the viscosity as
\begin{equation}
\label{viscosity}
\eta=r^{3}_{+}
\end{equation}
The relation (\ref{viscosity}) is identical to the viscosity of the hydrodynamical theory dual to Reissner-Nordstrom-AdS black branes (which is the case of infinite $\beta$) which we denote by $\eta_{\beta=\infty}$. Nonetheless, its functional dependence on temperature and chemical potential is different. For example, we can see this by simply expanding in $1/\beta^{2}$, upon which up to first order, we have
\begin{equation}
\label{eta}
\eta = \Bigg( \frac{\pi T}{2} \bigg(1+\sqrt{1+\frac{2\mu^{2}}{3g^{2}\pi^{2}T^{2}}}\,\bigg) \Bigg)^{3} - \frac{1}{\beta^{2}} \mathcal{G}\bigg(r_{+}, \mu_{0}, T_{0} \bigg)=\eta_{\beta=\infty} - \frac{1}{\beta^{2}} \mathcal{G}\bigg(r_{+}, \mu_{0}, T_{0} \bigg)
\end{equation}
where $T_{0}$ and $\mu_{0}$ denote the temperature and chemical potential of the infinite $\beta$ case, and $\mathcal{G}$ is a non-vanishing function. This is not surprising, in view of (\ref{temp}) and (\ref{chem}). Let us now consider the entropy density $s$ which can be computed by taking the partial derivative of $2m$ with respect to $T$ at fixed $\mu$:
\begin{equation}
\label{entropys}
s=\frac{\partial(2m)}{\partial T} \vert_{\mu}=\frac{\partial_{r_{+}}(2m)+\partial_{q}(2m)\frac{{\rm d}q}{{\rm d}r_{+}}}{\partial_{r_{+}}T+\partial_{q}T\frac{{\rm d}q}{{\rm d}r_{+}}}=4\pi r^{3}_{+}
\end{equation}
We observe that (\ref{entropys}) is actually equivalent to the Bekenstein-Hawking entropy density of the EBI black branes as calculated in \cite{Pang}. Taking the ratio of (\ref{viscosity}) to (\ref{entropys}) yields the universal ratio:
\begin{equation}
\label{ratio}
\frac{\eta}{s} = \frac{1}{4\pi}
\end{equation}
Note that both (\ref{viscosity}), (\ref{entropys}) and hence (\ref{ratio}) are results exact to all orders of $1/\beta$ even though we are working in a first-order derivative expansion in the metric and gauge fields. This is thus an improvement of the conclusion made in \cite{Cai} in which (\ref{ratio}) was calculated to first order in $1/\beta^{2}$ by using the Kubo formula to analyze gravitational perturbations of the same EBI black branes.

\section{Conclusion}

To summarize, we have begun with a class of boosted 5D black branes which are exact solutions to the EBI action of (\ref{action}), uplifted their temperature and boost parameters to be tranverse-coordinates-dependent, and then used a derivative expansion to construct the energy momentum tensor of the hydrodynamical theory living at the boundary. This technique - first introduced in the seminal work of \cite{Minwalla}, allowed us to deduce $\eta/s = 1/4\pi$ to all orders of inverse $\beta$ instead of only to first order of $1/\beta^{2}$ in \cite{Cai}. A brief re-look at our calculation will show that the crucial step lies in obtaining the coefficient of the $1/r^{4}$ term in one of the metric components. That $\alpha_{ij}$ is integrable is the key to the advantage of this method over the Kubo formula for this particular problem. Our result is an improvement of the conclusion in \cite{Cai}, and thus strengthens the perspective stated in \cite{Cai}: that quantum corrections to this ratio seem to arise mainly due to gravitational degrees of freedom.

A natural future direction is to construct and analyze the first-order charge current $J^{\mu}$ in greater details than that outlined in the Appendix, upon which we can read off the thermal and electrical conductivities, at least to certain orders of inverse $\beta$. One may then use the resulting correction to obtain at least a qualitative picture of how the nonlinearity and well-known screening effect \cite{Born} induced by the Born-Infeld term manifests itself in a hydrodynamical theory through, for example, corrections to the Wiedemann-Franz law \cite{Lifshitz} or other transport coefficients. As a gauge theory containing Born-Infeld terms is well-motivated by string theory \cite{Tseytlin}, it would be worthwhile to investigate further how much more the fluid/gravity correspondence can tell us about the hydrodynamical regime of such a theory.

\acknowledgments
The author is indebted to Ori Ganor and Petr Ho\v{r}ava for wonderful encouragements and advice.
\bigskip
\bigskip
\appendix
\section{Calculation of the metric and gauge correction terms $j_{i}$ and $a_{i}$}

In the following, we write down explicitly an outline for the calculation of the metric and gauge correction terms $j_{i}$ and $a_{i}$ to first order in $1/\beta^{2}$. Firstly, it is straightforward to decouple (\ref{coupled}) to obtain:
\begin{equation}
\label{A1}
\partial_{r}^{2}j_{i} - \frac{3}{r} \partial_{r}j_{i} - \frac{12q^{2}}{r^{8}f(r)}j_{i} = \mathcal{S}_{i}(r)
\end{equation}
\begin{equation}
\label{A2}
\partial_{r}a_{i} = \frac{2g^{2}}{\Theta \Upsilon} \bigg[ \frac{r}{2}\partial_{r}\bigg(\frac{\partial_{r}j_{i}}{r^{3}}\bigg) + \frac{3}{2r} \partial_{v}b_{i} \bigg]
\end{equation}
where we have defined
\begin{equation}
\label{A3}
\mathcal{S}_{i}=-\frac{12q^{2}}{r^{4}f} \frac{j_{i}(r_{+})}{r^{4}_{+}} - 3r\partial_{v}b_{i}+\frac{\sqrt{12}q}{gr^{4}f} \int^{r}_{r_{+}} \textrm{d}x\, x S_{i} \equiv \mathcal{S}_{(0)} + \frac{1}{\beta^{2}} \tilde{S}_{i} + \mathcal{O}(1/\beta^{4})
\end{equation}
In the infinite $\beta$ case, equation (\ref{A1}) admits an exact homogeneous solution, and the general solution for $j_{i}$ and $a_{i}$ can be found using method of variation of parameters. This has been done in \cite{Hur} and \cite{Banerjee}, with the resulting expressions already quite involved. In our case, we were not being able to find the homogeneous solution mainly due to the fact that $f(r)$ contains a hypergeometric function. This may imply that we have to resort to perturbation methods to solve for the two terms (once $j_{i}$ is obtained, an integration can be carried out to obtain $a_{i}$), and thus the reason for introducing $\tilde{S}_{i}$. In the following, we will write down explicitly an outline of how one can obtain $j_{i}$ and $a_{i}$ to first order in $1/\beta^{2}$. The procedure is very standard with the only subtle point being the $\beta$ dependence of $m$.
Thus, we expand:
\begin{equation}
\label{A4}
j_{i} = j^{(0)}_{i} + \bigg(\frac{1}{\beta^{2}} \bigg) \tilde{J}_{i}
\end{equation}
Now, in \cite{Hur} and \cite{Banerjee}, $j^{(0)}_{i}$ was computed to be
\begin{eqnarray}
\label{A5}
j^{(0)}_{i}&=&-r^{4}f_{0}(r) \int_{r}^{\infty} \textrm{d}x \zeta_{i}(x) f_{0}(x)\,x \int_{x}^{\infty} \frac{\textrm{d} y}{y^{5}f^{2}_{0}(y)}\cr
&&+ r^{4}f_{0}(r) \bigg( \int_{r}^{\infty} \frac{\textrm{d}x}{x^{5}f^{2}_{0}(x)} \bigg) \bigg( r^{3} \partial_{v} b_{i} + \int_{r}^{\infty} \textrm{d}x \big[ x f_{0}(x) \zeta_{i}(x)+3x^{2} \partial_{v}b_{i} \big] \bigg)
\end{eqnarray}
where
\begin{equation}
\label{A6}
\zeta_{i}(r) = -3r\partial_{v}b_{i} - \frac{2\sqrt{3}q}{r^{4}f_{0}(r)} \bigg(2\sqrt{3}q\frac{j^{(0)}_{i} (r_{+})}{r^{4}_{+}} - \sqrt{3} \bigg( \frac{1}{r} - \frac{1}{r_{+}} \bigg) \big( \partial_{i}q + q\partial_{v}\beta_{i} \big) + \big(r - r_{+} \big) \frac{e}{g} F^{\rm{ext}}_{vi} \bigg)\,\\
\end{equation}
\begin{equation}
\label{A7}
f_{0}(r) = 1 - \frac{2m_{0}}{r^{4}} + \frac{q^{2}}{r^{6}},\,\qquad m_{0} \equiv \frac{1}{2} \left( r_{+}^{4} + \frac{q^{2}}{r_{+}^{2}} \right)
\end{equation}
One notes carefully that
\begin{equation}
\label{A8}
f(r)\approx f_{0} \bigg(1 + \frac{1}{\beta^{2}} \frac{\tilde{c}}{f_{0}} \bigg)\,,\qquad \tilde{c}= \frac{3g^{2}q^{4}}{4r^{4}}\big(r^{-8}_{+} - r^{-8}\big)
\end{equation}
Also, it is straightforward to compute $\tilde{S}$ to be
\begin{eqnarray}
\label{A9}
\tilde{S} &=& -\frac{-12q^2}{r^{4}f_{0}}\Bigg( \frac{\tilde{J}(r_{+})}{r^{4}_{+}} - \frac{\tilde{c}j_{0}(r_{+})}{f_{0}r^{4}_{+}}\Bigg)\cr
&&+\frac{\sqrt{12}q}{gr^{4}f_{0}} \Bigg( 12g^{3}q^{2} \bigg[ \bigg(\frac{5}{\sqrt{12}} \partial_{i}q + \frac{21\sqrt{3}}{8}q\partial_{v}b_{i} \bigg) \frac{1}{7}\big(r^{-7}-r^{-7}_{+} \big) - \frac{e}{2g}F^{\rm{ext}}_{vi} \big(r^{-5} - r^{-5}_{+} \big) \bigg] \cr
&&-\frac{g\tilde{c}}{f_{0}} \bigg[ \sqrt{3} \big( \partial_{i}q+q\partial_{v}b_{i}\big)\big(r^{-1}-r^{-1}_{+}\big) -\frac{e}{g}F^{\rm{ext}}_{vi} (r - r_{+}) \bigg] \Bigg)
\end{eqnarray}
Substituting (\ref{A8}) into (\ref{A1}), we arrive at an equation for $\tilde{J}_{i}$ to be
\begin{equation}
\label{A10}
\partial_{r}^{2}\tilde{J}_{i} - \frac{3}{r} \partial_{r}\tilde{J}_{i} - \frac{12q^{2}}{r^{8}f_{0}(r)}\tilde{J}_{i} = \tilde{S}_{i}(r) - \frac{12q^{2}\tilde{c}j_{0}(r)}{r^{8}f^{2}_{0}(r)} \equiv \mathcal{F}_{i}(r)
\end{equation}
Then by the method of variation of parameters, we obtain a closed-form expression for $\tilde{J}$ to be
\begin{equation}
\label{A11}
\tilde{J}_{i} = -r^{4}f_{0}(r) \int^{\infty}_{r} \textrm{d}x x f_{0}(x) \mathcal{F}_{i}(x) \int^{\infty}_{x} \frac{\textrm{d}y}{y^{5}f^{2}_{0}} + r^{4}f_{0}(r) \int^{\infty}_{r} \frac{\textrm{d}x}{x^{5}f^{2}_{0}} \int^{\infty}_{r} \textrm{d} x x f_{0}(x) \mathcal{F}_{i}(x)
\end{equation}
Clearly from (\ref{A2}), another integration will yield $a_{i}$ to first order in $1/\beta^{2}$.
The exact expressions for $j_{i}$ and $a_{i}$ are not needed for the computation of the conserved current $J^{\mu}$,
but only their asymptotic behaviors. In particular, the coefficient of the term in $1/r^{2}$ in $a_{i}$ will give $J^{\mu}$.
However, to write it covariantly and extract first-order linear transport coefficients - such as thermal/electrical conductivities from it, we will need to take the $r \rightarrow r_{+}$ limit of (\ref{A11}) due to the $\tilde{J}(r_{+})$-term in (\ref{A9}).
Although this could be achieved in the infinite $\beta$ problem, the integral form of $j_{0}$ might imply that
we need some numerics to calculate/estimate this limit. We leave this and related applications to future work.

\bigskip\bigskip

\end{document}